\documentclass[epj-spec]{svjour}
\usepackage{graphicx}

\begin{document}

\title{Describing Charmonium Correlation Functions in Euclidean Time}
\author{\'Agnes M\'ocsy\inst{1}\fnmsep \thanks{\email{mocsy@quark.phy.bnl.gov}} \and
P\'eter Petreczky\inst{2}\fnmsep\thanks{\email{petreczk@bnl.gov}}}
\institute{RIKEN-BNL Research Center, Brookhaven National Laboratory, Upton, NY 11973 USA \and
Physics Department and RIKEN-BNL Research Center, Brookhaven National Laboratory, Upton, NY 11973 USA}
\abstract{
We present a detailed investigation of the 
quark mass-dependence  
of  charmonium correlators in Euclidean-time obtained using a  
potential model, as well as the comparison with results on 
isotropic lattice calculations performed at several lattice spacings.
} 
%
\maketitle  
\section{Introduction}
\label{intro}

It was argued long ago that melting of quarkonia
above the deconfinement transition can serve as a 
signature of quark gluon plasma formation in heavy
ion collisions \cite{MS86}. The basic idea behind this
proposal was that due to color screening the potential 
between quark and anti-quark will not provide sufficient
binding at high temperature. This problem can be formulated
more rigoruosly in terms of quarkonium spectral functions, which can be, 
in principle, extracted from Euclidean-time meson correlation functions calculated on the lattice.
Attempts doing this based on the Maximum Entropy Method 
(MEM) have been discussed over the last few
years. The initial interpretation of data led to the conclusion that the $1S$ charmonia states
survive in the deconfined medium up to temperatures of about
$1.6T_c$, with $T_c$ being the transition temperature 
\cite{umeda02,asakawa04,datta04,swan}. Recent analysis, however,
has shown that, although  MEM can be used to extract reliably  
quarkonium spectral functions at zero temperature, at finite 
temperature it has sever limitations \cite{jako07}.

In a recent study we explored quarkonium correlators in Euclidean
time using a potential model \cite{mocsy07}. 
We have shown that the temperature (in)dependence 
of quarkonium correlators can be explained provided that color
screening melts most of the quarkonium states. The absence of bound states
in a quarkonium spectral function is compensated by a large threshold
enhancement, leaving the Euclidean correlators unchanged \cite{mocsy07}. 
This analysis done in QCD with only heavy quarks has been extended
to 2+1 flavor QCD, and was used to estimate the upper limit on the dissociation temperatures of
the different quarkonium states \cite{mocsy07a}.
The comparison of correlators at zero temperature was 
done using results from calculations done on isotropic lattices. 
This is  because only for isotropic lattices the renormalization constants for the local
meson currents are known (see discussion in Ref. \cite{datta04}).
The temperature-dependence of the correlators calculated in 
potential models has been compared against the results from
anisotropic lattices. In this paper we extend our previous studies
by comparing against  isotropic lattice calculations at smaller
lattice spacings. While in our previous calculations we had 
to introduce a prefactor when comparing lattice data with potential
model predictions, here we do a parameter-free comparison as it
is possible to use the prefactor's previously determined value.

\section{Charmonium Spectral Functions in Potential Model}

Since the seminal paper by Matsui and Satz the problem of charmonium
dissolution has been studied in potential models 
\cite{karsch88,ropke88,hashimoto88,digal01,wong04,mocsy05,mocsy06,rapp06,alberico07,wong06}. 
While the early  studies used phenomenological potential more recent studies rely on lattice
calculations of the static quark anti-quark free energy. Recently attempts to calculate
quarkonium properties at finite temperature using resummed perturbation theory have been made
\cite{laine06,laine07}. 
Free energy calculations are done in pure gluodynamics, 3-flavor and 2-flavor  
QCD \cite{okacz,petrov04,okacz05}, and preliminary results are also available in 
the physically relevant case of one heavy strange quark 
and two light quarks  \cite{kostya} (quark masses correspond to pion mass of about $200$~MeV).
Since the lattice calculations of the spectral functions have severe limitations, 
in \cite{mocsy05,mocsy06} it has been pointed out, that comparison between the lattice data 
should be done at the level
of the Euclidean time correlators, for which the numerical
results are much more reliable. Recent studies 
following this line have also been presented in Refs. \cite{rapp06,alberico07,{wong06}}. 
However, for such comparison to be meaningful lattice artifacts in the Euclidean
correlators has to be understood. Therefore, it is important to do the comparison
of the potential model results with lattice calculations performed at several
lattice spacings.

For heavy quarks (here we only consider charm) the spectral function can be related to the non-relativistic Green's functions 
\begin{eqnarray}
&
\displaystyle
\sigma(\omega)=K \frac{6}{\pi} {\rm Im} G^{nr}(\vec{r},\vec{r'},E)|_{\vec{r}=\vec{
r'}=0}\, ,\\[2mm]
&
\displaystyle
\sigma(\omega)=K \frac{6}{\pi}\frac{1}{m_c^2} {\rm Im} \vec{\nabla} 
\cdot \vec{\nabla'} G^{nr}(\vec{r},\vec{r'},E)|_{\vec{r}=\vec{r'}=0}\, ,
\label{green_sc} 
\end{eqnarray}
for $S$-wave, and $P$-wave charmonia, respectively. Here $E=\omega-2 m_c~$.
At leading order $K=1$. Relativistic and higher order perturbative corrections
will lead to a value different from unity.
The non-relativistic Green's function satisfies the Schr\"odinger equation
\begin{equation}
\left [ -\frac{1}{m_c} \vec{\nabla}^2+V(r)-E \right ] G^{nr}(\vec{r},\vec{r'},E) = \delta^3(r-r')\, .
\label{schroedinger}
\end{equation}
The numerical method for solving this equation is presented in \cite{mocsy07}.
At zero temperature we use the Cornel potential $V(r)=-\alpha/r+\sigma r$ with parameters 
motivated by lattice results on static potential : $\alpha=\pi/12$ and 
$\sigma=(1.65-\pi/12)r_0^{-2}$ (see Ref. \cite{mocsy07} for further details.).   
At finite temperature we use a potential motivated by lattice results on the
singlet free energy of static quark anti-qiark pair and which is defined in section 
IV of Ref. \cite{mocsy07}.
At large energies, away from the threshold, the non-relativistic treatment is  
not applicable. 
The spectral function in this domain, however,  can be calculated using perturbation theory. 
As in our previous work, we smoothly match the non-relativistic calculation of the 
spectral function to the relativistic perturbative result \cite{mocsy07}.  
Euclidean time correlators $G(\tau,T)$ at some temperature $T$ can 
be calculated from the spectral functions
using the integral representation
\begin{equation}
G_{rec}(\tau,T)=\int_0^{\infty} d \omega \sigma(\omega,T) K(\omega,\tau,T)\, .
\label{spect_rep}
\end{equation}
Here the integration kernel is 
\begin{equation}
\displaystyle
K(\omega,\tau,T)=\frac{\cosh \omega (\tau-1/(2T))}{\sinh \left( \omega/(2T) \right)}\, .
\label{kernel_T}
\end{equation}

\section{Correlators at Zero Temperature}

In this section we discuss the comparison of the model calculations with zero temperature
lattice data from isotropic lattices \cite{datta04,datta_tbp}. The lattice spacing has been
fixed using the Sommer-scale $r_0=0.5$fm. Its value is slightly larger than the one used in Ref. 
\cite{datta04}, since there the string tension of $\sqrt{\sigma}=420~$MeV has been 
used to set the scale. Calculations have
been done at several values of the charm quark mass, but unfortunately none of them exactly at the physical value.
The renormalization constants of the lattice operators has been calculated in 1-loop tadpole
improved perturbation theory (see Ref. \cite{datta04} for further details). 
In Table \ref{tab:values} we give the value of the gauge coupling $\beta=6/g^2$ 
used in lattice calculations, 
the corresponding lattice spacings, the renormalization constants $Z_V$ and $Z_P$ for vector and the
pseudo-scalar current,  
as well as the estimated masses of the ground state $\eta_c$ meson. 
The values of the quark masses used in potential model calculations are
given here as well.  All the lattice calculations have been done on a $48^3 \times 24$ 
lattice. 
This corresponds to $0.6T_c$ for the larger lattice spacing, and $0.75T_c$ for the smaller lattice spacing. 
This is the reason why the mass of the $\eta_c$
has a large uncertainty. For the interested reader's convenience, in Table \ref{tab:values}  
we also provide the parameters used in the analysis of Ref. \cite{mocsy07}. 
\begin{table}
\begin{center}
\begin{tabular}{|c|c|c|c|c|c|}
\hline
$\beta$   &   a [fm]   &  $Z_V$ & $Z_P$ & $M_{\eta_c}$ [GeV] &  $m_c$ [GeV] \\
\hline
6.499     &   0.0451   &  0.975 & 0.847 &  2.622(50)        &     0.95 \\
          &   0.0451   &  1.040 & 0.904 &  3.271(50)        &     1.34\\
          &   0.0451   &  1.124 & 1.032 &  4.495(10)        &     2.00\\
\hline
6.640     &   0.0377   &  1.007 & 0.881 &  3.297(270)       &     1.34\\
\hline
7.192     &   0.0170   &  0.936 & 0.839 &  4.023(52)        &     1.70\\
\hline
\end{tabular}
\caption{Lattice parameters of the correlators used in the present analysis, 
the masses of the $\eta_c$ obtained on isotropic lattices, 
and the corresponding charm quark masses used in the potential model.}
\label{tab:values}
\end{center}
\end{table}  

\begin{figure}
\begin{center}
\includegraphics[width=8.5cm]{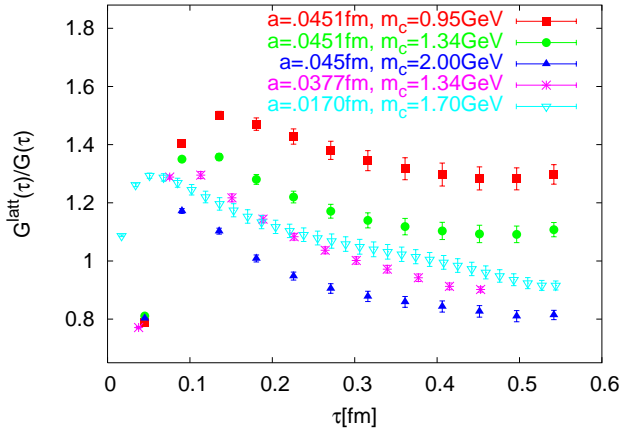}
\caption{The ratio of the pseudo-scalar correlators calculated on the lattice
to the ones calculated in our model for different quark masses and
lattice spacings.}
\label{fig:corrps}
\end{center}
\end{figure}

We used the previously determined values of the $K$ factors: $K=2.0$ for
scalar and pseudo-scalar channels and $K=0.8$ for the vector channel \cite{mocsy07}. 
In Fig. \ref{fig:corrps} we show the ratio of the pseudo-scalar correlators
calculated on isotropic lattices for the parameters mentioned above and the correlator
calculated in our model. Here we also show the results of the calculations at heavier
quark mass $m_c=1.7$GeV considered in Ref. \cite{mocsy07}. For the smallest two lattice
spacings we find a reasonable agreement for this ratio. For the coarser lattice, 
$a=0.0451$fm our model does not seem to dscribe the mass dependence of the correlator very well.
This could be due to the quark mass dependence of the $K$ factor and/or lattice artifacts. 
Note that for $m_c=1.34$GeV there is about $20\%$ discrepancy between the results obtained
at two different lattice spacings. By some tuning of the $K$ factors, namely
choosing $K=2.4$ and $2.0$ for $m_c=0.95$GeV a $m_c=2.0$GeV a much better agreement
between different lattice data can be obtained. 

\begin{figure}                 
\begin{center}                                                              
\includegraphics[width=8.5cm]{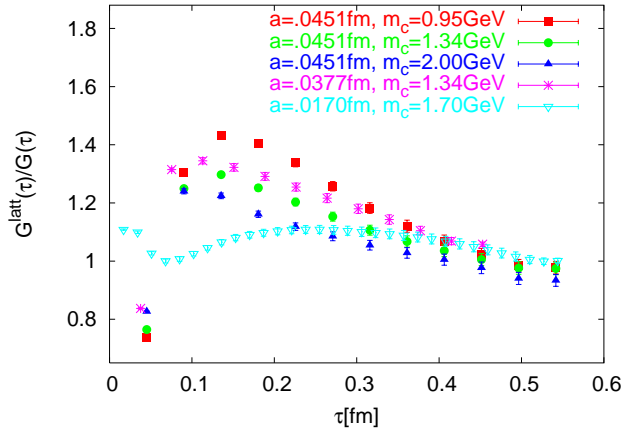}                                             
\caption{The ratio of the vector correlators calculated on the lattice                            
to the ones calculated in our model for different quark masses and                                    
lattice spacings.}                                                                                     
\label{fig:corrvc}     
\end{center}                                                                               
\end{figure}

We also considered the vector correlators. In Fig. \ref{fig:corrvc} we show
the ratio of the lattice data to our model predictions. As one can see from
the figure the model can capture the quark mass depence of the correlators
calculated on the lattice much better in this case at least for $\tau>0.3$fm.
At smaller Euclidean times we see significant deviations of this ratio from unity
which is presumably due to lattice artifatcs. 

\section{Temperature-dependence of Pseudo-scalar Correlators}

In this section we study the temperature-dependence of the pseudoscalar
correlator for the different quark masses considered also in the previous Section.
For comparison we again use the lattice data from isotropic lattice simulations
\cite{datta04,datta_tbp}. As customary, in order to eliminate the trivial temperature dependence
in the correlators we consider the ratio $G(\tau,T)/G_{rec}(\tau,T)$, with 
\begin{equation}                                                                                             
G_{rec}(\tau,T)=\int_0^{\infty} d \omega \sigma(\omega,T=0) K(\omega,\tau,T)\, .        
\label{grec} 
\end{equation}

\begin{figure}    
\begin{center}                                                          
\includegraphics[width=8.5cm]{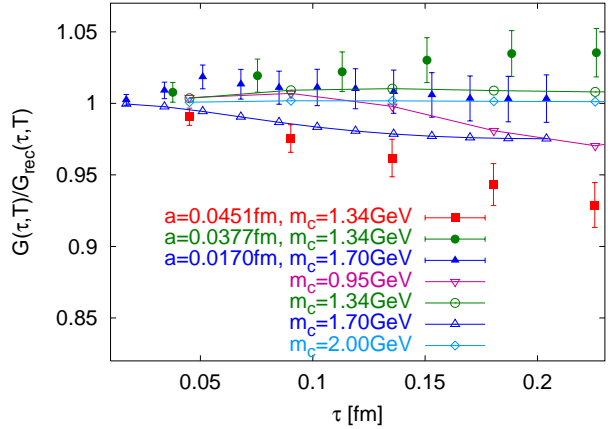}                                             
\caption{The ratio $G(\tau,T)/G_{rec}(\tau,T)$ calculated for different
lattice spacings at $1.5T_c$.}                                                                               
\label{fig:rat}     
\end{center}                                                                                   
\end{figure}

The lattice data for this ratio together with potential model calculations
is shown in Fig. \ref{fig:rat} for several values of the lattice spacings
at temperature $1.5T_c$.
The lattice calcuations have been performed on $48^3 \times 10$, $48^3 \times 12$
and $64^3 \times 24$ lattices for $a=0.0451$fm, $a=0.0377$fm and $a=0.0170$fm respectively.
The potential model predicts the ratio $G(\tau,T)/G_{rec}(\tau,T)$ to be close
to unity for all quark masses around the charm quark mass, in agreement with
the analysis donein Ref. \cite{mocsy07}. We find that the mass dependence of
$G/G_{rec}$ is about few percent.
One can see from the figure that correlators calculated at lattice spacing
$a=0.0451~$fm and $a=0.0377~$fm are quite different, 
even though the quark mass in
these calculations is  the same. From this we conclude that lattice artifacts are significant, 
and thus care is needed when comparing the temperature-dependence of the correlators on coarse
lattices with model calculations. 

The temperature dependence of the charmonium correlators is more pronounced
in other channels and is mostly due to the zero mode contribution \cite{mocsy06,derek,umeda07,alberico_new}.

\section{Conclusions}

In this paper we have studied charmonium correlators
at zero and at finite temperatures in potential model. We analyzed the model for several quark masses
near the physical charm quark mass, and compared the results
with available lattice data from isotropic lattices. At zero temperature we 
found a reasonably good agreement between the model calculations and the lattice
data using the $K$ factor fixed in the previous calculations done for 
$M_{\eta_c}=4.023$GeV \cite{mocsy07}.
We find that $G/G_{rec}$ is close to unity and shows only a weak dependence on
the quark mass in the quark mass region around the charm mass.  

\section*{Acknowledgments}
We extend our special thank to P.~L\'evai, T.~Bir\'o and T.~Cs\"org\"o for inviting us to, and for providing very kind hospitality during the Workshop in memoriam Prof. J.~Zim\'anyi. We further thank P. ~Sorensen for careful reading of the manuscript and valuable comments.
We are also grateful to S. ~Datta for providing unpublished lattice data on charmonium
correlators. 
This work has been supported by U.S. Department of Energy under Contract No. DE-AC02-98CH10886.


\begin{thebibliography}{99}

\bibitem{MS86}
T.~Matsui and H.~Satz,
Phys.\ Lett.\ B {\bf 178}, 416 (1986).

\bibitem{umeda02}
T.~Umeda, K.~Nomura and H.~Matsufuru,
Eur.\ Phys.\ J.  C {\bf 39S1}, 9 (2005)
[arXiv:hep-lat/0211003].

\bibitem{asakawa04}
M.~Asakawa and T.~Hatsuda,
Phys.\ Rev.\ Lett.\  {\bf 92}, 012001 (2004)

\bibitem{datta04}
S.~Datta, F.~Karsch, P.~Petreczky and I.~Wetzorke,
Phys.\ Rev.\ D {\bf 69}, 094507 (2004)

\bibitem{swan}
G.~Aarts et al,
arXiv:hep-lat/0610065; 
  Nucl.\ Phys.\  A {\bf 785}, 198 (2007)
  [arXiv:hep-lat/0608009]; arXiv:0705.2198 [hep-lat]

\bibitem{jako07}
  A.~Jakovac, P.~Petreczky, K.~Petrov and A.~Velytsky,
  Phys.\ Rev.\  D {\bf 75}, 014506 (2007)  

\bibitem{mocsy07}
  \'A.~M\'ocsy and P.~Petreczky,
  arXiv:0705.2559 [hep-ph].

\bibitem{mocsy07a}
  \'A.~M\'ocsy and P.~Petreczky,
  arXiv:0706.2183 [hep-ph].

\bibitem{datta_tbp}
S. Datta, P. Petreczky, to be published

\bibitem{karsch88}
F.~Karsch, et al.,
Z.\ Phys.\ C {\bf 37}, 617 (1988);

\bibitem{ropke88}
G. R{\"o}pke, D. Blaschke, H. Schulz, Phys.\ Rev.\ D {\bf 38}, 3589 (1988)

\bibitem{hashimoto88}
T.  Hashimoto et al., Z.\ Phys.\ C {\bf 38}, 251 (1988)

\bibitem{digal01}
S.~Digal, et al.,
Phys.\ Lett.\ B {\bf 514}, 57 (2001);
Phys.\ Rev.\ D {\bf 64}, 094015 (2001);

\bibitem{wong04}
  C.~Y.~Wong,
  Phys.\ Rev.\ C {\bf 72}, 034906 (2005);

\bibitem{mocsy05}                                                              
  \'A.~M\'ocsy and P.~Petreczky,                                               
  Eur.\ Phys.\ J.\ C {\bf 43}, 77 (2005)                                       

\bibitem{mocsy06}                                                              
  \'A.~M\'ocsy and P.~Petreczky,                                               
  Phys.\ Rev.\ D {\bf 73}, 074007 (2006)                                       

\bibitem{rapp06}
  D.~Cabrera, R.~Rapp,
  arXiv:hep-ph/0610254.

\bibitem{alberico07}
  W.~M.~Alberico, A.~Beraudo, A.~De Pace and A.~Molinari,
  Phys.\ Rev.\  D {\bf 75}, 074009 (2007)

  \bibitem{wong06}
  C.~Y.~Wong and H.~W.~Crater,
  Phys.\ Rev.\  D {\bf 75}, 034505 (2007)
  [arXiv:hep-ph/0610440].

\bibitem{laine06}
  M.~Laine, O.~Philipsen, P.~Romatschke and M.~Tassler,
  JHEP {\bf 0703}, 054 (2007)
  [arXiv:hep-ph/0611300].

\bibitem{laine07}
  M.~Laine,
  arXiv:0704.1720

\bibitem{okacz}
  O.~Kaczmarek, et al.,
  Phys.\ Lett.\  B {\bf 543}, 41 (2002); 
  Phys.\ Rev.\  D {\bf 70}, 074505 (2004)
  [Erratum-ibid.\  D {\bf 72}, 059903 (2005)]

\bibitem{petrov04}
  P.~Petreczky, K.~Petrov,
   Phys.\ Rev.\  D {\bf 70}, 054503 (2004)

\bibitem{okacz05}
  O.~Kaczmarek, F.~Zantow,
  Phys.\ Rev.\  D {\bf 71}, 114510 (2005)

\bibitem{kostya}
  K.~Petrov  [RBC-Bielefeld Coll.],
  arXiv:hep-lat/0610041.

\bibitem{derek}
  P.~Petreczky and D.~Teaney,
  Phys.\ Rev.\  D {\bf 73}, 014508 (2006)
  [arXiv:hep-ph/0507318].

\bibitem{umeda07}
  T.~Umeda,
  arXiv:hep-lat/0701005.

\bibitem{alberico_new}
W.M. Alberico et. al, arXiv:07062846

\end{thebibliography}
\end{document}